\documentclass[aps,
prr,
twocolumn,
superscriptaddress,
dvipsnames,
longbibliography,]{revtex4-1}

\usepackage{amssymb,amsfonts,amsmath} 
\usepackage{graphicx,epsfig,psfrag}% Include figure files 
\usepackage{color}
\usepackage{natbib}
\usepackage{url}
\usepackage[breaklinks=true]{hyperref}
\usepackage{mathtools}
\usepackage{subfigure}
\hypersetup{
        colorlinks = true,
        urlcolor = blue,
        citecolor = blue
}
\usepackage{bookmark}
\usepackage{epic}
\usepackage{longtable}
\usepackage{bbm}
\usepackage{ulem}
\usepackage{braket}
\usepackage{xcolor}
\usepackage{csquotes}
\usepackage{tikz}
\usepackage{extarrows}
\usetikzlibrary{trees}
\usetikzlibrary{decorations.pathmorphing}
\usetikzlibrary{decorations.markings}
\usetikzlibrary{positioning,arrows}
\usetikzlibrary{shapes.arrows}
\usepackage{leftidx}
\usepackage{framed}
\usepackage{tcolorbox}
\usepackage{bm}
\usepackage{physics}
\usepackage{lipsum}

\tikzset{->-/.style={decoration={
  markings,
  mark=at position #1 with {\arrow{>}}},postaction={decorate}}}

\newcommand\cc[1][]{c_{#1}}
\newcommand\ccdag[1][]{c^\dagger_{#1}}

\definecolor{lime}{HTML}{A6CE39}
\DeclareRobustCommand{\orcidicon}{%
    \begin{tikzpicture}
    \draw[lime, fill=lime] (0,0) 
    circle [radius=0.16] 
    node[white] {{\fontfamily{qag}\selectfont \tiny ID}};
    \draw[white, fill=white] (-0.0625,0.095) 
    circle [radius=0.007];
    \end{tikzpicture}
    \hspace{-2mm}
}

\foreach \x in {A, ..., Z}{%
    \expandafter\xdef\csname orcid\x\endcsname{\noexpand\href{https://orcid.org/\csname orcidauthor\x\endcsname}{\noexpand\orcidicon}}
}
\newcommand{\orcid}[1]{\href{https://orcid.org/#1}{\textcolor[HTML]{A6CE39}{\orcidicon}}}

\begin{document}

\title{Many-body localization in the interpolating Aubry-Andr\'{e}-Fibonacci model}

\author{Antonio \v{S}trkalj \orcid{0000-0002-9062-6001}}
\email{as3157@cam.ac.uk}
\affiliation{\mbox{T.C.M. Group, Cavendish Laboratory, J. J. Thomson Avenue,
University of Cambridge, Cambridge, CB3 0HE, UK}}
\affiliation{Institute for Theoretical Physics, ETH Z\"{u}rich, 8093 Z\"{u}rich, Switzerland}

\author{Elmer V. H. Doggen}
\affiliation{\mbox{Institute for Quantum Materials and Technologies, Karlsruhe Institute of Technology, 76021 Karlsruhe, Germany}}
\affiliation{\mbox{Institut f\"{u}r Theorie der Kondensierten Materie, Karlsruhe Institute of Technology, 76128 Karlsruhe, Germany}}

\author{Igor V. Gornyi}
\affiliation{\mbox{Institute for Quantum Materials and Technologies, Karlsruhe Institute of Technology, 76021 Karlsruhe, Germany}}
\affiliation{\mbox{Institut f\"{u}r Theorie der Kondensierten Materie, Karlsruhe Institute of Technology, 76128 Karlsruhe, Germany}}
\affiliation{Ioffe Institute, 194021 
St.~Petersburg, Russia}

\author{Oded Zilberberg}
\affiliation{Institute for Theoretical Physics, ETH Z\"{u}rich, 8093 Z\"{u}rich, Switzerland}

%==========================================================
\begin{abstract} 
We investigate the localization properties of a spin chain with an antiferromagnetic nearest-neighbour coupling, subject to an external quasiperiodic on-site magnetic field. The quasiperiodic modulation interpolates between two paradigmatic models, namely the Aubry-Andr\'{e} and the Fibonacci models. We find that stronger many-body interactions extend the ergodic phase in the former, whereas they shrink it in the latter. Furthermore, the many-body localization transition points at the two limits of the interpolation appear to be continuously connected along the deformation of the quasiperiodic modulation. As a result, the position of the many-body localization transition depends on the interaction strength for an intermediate degree of deformation. Moreover, in the region of parameter space where the single-particle spectrum contains both localized and extended states, many-body interactions induce an anomalous effect: weak interactions localize the system, whereas stronger interactions enhance ergodicity. We map the model's localization phase diagram using the decay of the quenched spin imbalance in relatively long chains. This is accomplished employing a time-dependent variational approach applied to a matrix product state decomposition of the many-body state. Our model serves as a rich playground for testing many-body localization under tunable potentials. 
\end{abstract}

%==========================================================
\date{\today} 
\maketitle
%==========================================================

%=====================%
\section{Introduction}
The study of material properties most commonly begins by assuming a periodic crystalline structure within which extended Bloch waves manifest and form dispersive bands~\cite{Ashcroft1976}. Breaking the periodicity by disorder~\cite{anderson1958} or by incommensurate potentials~\cite{senechal1995} can localize the single-particle states of the system, leading to insulating behavior. Keeping the periodicity instead but introducing many-body interactions also breaks the Bloch picture and can lead to localization, e.g., through the formation of a charge density wave or a Mott insulator~\cite{Mott1949}. The fate of the system's conduction properties in the presence of both many-body effects and (quasi-)disorder is very rich and depends on numerous details.
Specifically, the physics of quantum many-body interacting systems can be strongly altered by the presence of disorder, which can drive them from an ergodic to a many-body localized (MBL) phase~\cite{altshuler1997, gornyi2005, basko2006, oganesyan2007, znidaric2008, pal2010, nandkishore2015, abanin2019}. The latter is interesting due to the fact that a closed MBL system does not thermalize at any time scale and remains robust to small perturbations, such as changing the interaction strength, the strength of disorder, and/or the temperature. Correspondingly, the dynamics of the system is frozen deep within the MBL phase, while close to the transition point between the ergodic and the MBL phase, power laws appear in transport properties. Systems exhibiting MBL are therefore interesting both from (i) the perspective of fundamental science, since they provide a test-bed for general mechanisms by which ergodicity is broken in quantum systems, and (ii) their technological applications, since they can host types of order that cannot be present in equilibrium, which can be used for information storage, and isolation of quantum information processing devices. 

Due to its complexity, the majority of works in the field of many-body localization consider one-dimensional systems, which are numerically more tractable. In one-dimensional systems, the single-particle spectrum is localized by randomly distributed disorder~\cite{anderson1958}, and the many-body interactions can delocalize the system through hybridization between the localized orbits or fail to do so such that an MBL phase forms~\cite{gornyi2005, basko2006, znidaric2008, pal2010}.  
Noninteracting quasiperiodic systems instead show richer localization phenomena in one dimension compared to randomly disordered systems. Depending on the specific model, they are known to host (i) a pure point localization transition at a finite threshold for the Aubry–Andr\'{e} (AA) model~\cite{aubry1980, harper1955}, (ii) critical states for the Fibonacci chain~\cite{kohmoto1983, ostlund1983}, as well as (iii) mobility edges~\cite{ganeshan2015, luschen2018, strkalj2020}. Interestingly, also for such quasiperiodic systems it was shown theoretically \cite{iyer2013, bera2017, lev2017, doggen2019, mace2019, varma2019, Chiaracane2021} and experimentally \cite{schreiber2015, luschen2017, luschen2018} MBL phases appear, but the mechanism that leads to its formation depends on the fine details of the model.

For example, the quasiperiodicity in the AA model arises from an on-site cosine modulation that is incommensurate with the underlying periodic lattice spacing. The Fibonacci model instead involves two discrete on-site values that interchange throughout the system according to the Fibonacci sequence. The noninteracting Aubry–Andr\'{e} model is known to have a metal-to-insulator transition at finite strength of quasiperiodic modulation simultaneously for all eigenstates~\cite{aubry1980, harper1955}, while the Fibonacci model always has critical eigenstates that are fractal~\cite{kohmoto1983, ostlund1983}. Adding many-body interactions to the AA model results in the shift of the localization transition in favour of an ergodic phase~\cite{doggen2019}, where transport is diffusive~\cite{Znidaric2018,Znidaric2021}.
In the Fibonacci model, instead, interactions destroy the fractal critical states and introduce an MBL phase transition~\cite{mace2019, Chiaracane2021}. Crucially, the full details behind the MBL transition in  interacting quasiperiodic models remain unknown. Hence, investigation of different interacting quasiperiodic models can provide us with more insight into the interplay between interaction and the models' exotic single-particle localization phenomena.

Interestingly, the AA and Fibonacci models can be viewed as two limits of the interpolating Aubry--Andr\'{e}-Fibonacci model~\cite{kraus2012} (IAAF). The experimental realization of this model was reported recently in Ref.~\cite{strkalj2020}, wherein some of us have explored the localization phase diagram of the noninteracting IAAF model, and revealed how the critical states of the Fibonacci model form alongside a cascade of localization-delocalization transitions as the model is tuned from the AA to the Fibonacci limit. This cascade of transitions occurs non-uniformly throughout the spectrum developing many non-trivial mobility edges in the Hermitian~\cite{strkalj2020} and non-Hermitian~\cite{Zhai2021} version of the model. Although the physics of the noninteracting model is now well understood, the model's many-body localization properties have thus far not been explored. Specifically, how does the rich localization physics of the single-particle IAAF model with its uniform and nonuniform metal-to-insulator transitions, critical states, and mobility edges, interplay with many-body interactions? More precisely, a single-particle spectrum that is not uniformly localized (mobility edges) is predicted to exhibit a peculiar interplay with many-body interactions tending to delocalization through the ergodic parts of the single-particle spectrum~\cite{Potter2014,Gopalakrishnan2014}.

In this work, we set out to investigate the localization phase diagram of the interacting IAAF (i-IAAF) model and answer two main questions: 
(i) How are the MBL transition points of the AA and Fibonacci models connected? 
(ii) What happens to the cascade of localization-delocalization transitions of the noninteracting model in the presence of interactions?. In the latter, we have regions in parameter space where a nontrivial mobility edge manifests in the noninteracting limit, with coexistence of extended and localized states in the system.  In this regime, we observe anomalous behavior with increasing interaction strength, where weak interactions localize the system, while stronger interactions again drive it to the ergodic phase. Our results provide a first glimpse into the complex many-body physics found in the i-IAAF model, opening the path for additional studies in this tunable setting.

The paper is outlined as follows: in Section~\ref{sec:model}, we introduce the Hamiltonian of the IAAF model and discuss the main localization properties of the single-particle model. In the same section, we also review the known results related to the interacting AA and Fibonacci models. In Section~\ref{sec:results}, the main results of this paper are presented; namely we (i) describe our results for the Fibonacci model at two different interactions strengths, (ii) obtain the many-body phase diagram of the i-IAAF model, and (iii) discover the anomalous impact of many-body interactions on the regime where the model exhibits mobility edges. 
In Section~\ref{sec:discussion}, we discuss in detail our results from Section~\ref{sec:results}, in particular, the physical mechanisms behind the many-body phase diagram of the i-IAAF model. Our findings are summarized in Section~\ref{Sec:Conclusion}.
Technical details are relegated to Appendices.

%=====================%
\section{Model   \label{sec:model}}
We consider a finite spin-$1/2$ chain with open boundary conditions containing $L$ sites in a quasiperiodic magnetic field. The Hamiltonian of the system is given by 
\begin{align}
    \resizebox{.90\hsize}{!}{$\displaystyle{H = \sum_{j=1}^{L}\Big[ J (S_j^x S_{j+1}^x + S_j^y S_{j+1}^y ) + \Delta S_j^z S_{j+1}^z  + h_j S_j^z \Big],}$}
    \label{eq:XXZ_hamiltonian}
\end{align}
where $S_j^{x,y,z}$ denote standard Pauli matrices, the in-plane coupling amplitude $J=1$ sets the energy scale of the problem, and we work in units with $\hbar=1$. In this parametrization, $\Delta=1$ corresponds to the isotropic Heisenberg chain. 
The chain is subjected to a spatially-modulated magnetic field
$h_j = \lambda V_j(\beta)$  of strength $\lambda\geq 0$, where the modulation function $V_j(\beta)$ is defined as
\begin{align}
    V_j (\beta) = -\frac {\tanh {[\beta \cos{(2\pi b j + \phi)} -\beta 
    \cos{(\pi b)} }]} {\tanh \beta} \, .
    \label{eq:potential}
\end{align}

\begin{figure}[t!]
	\centering
    \includegraphics[width=\columnwidth]{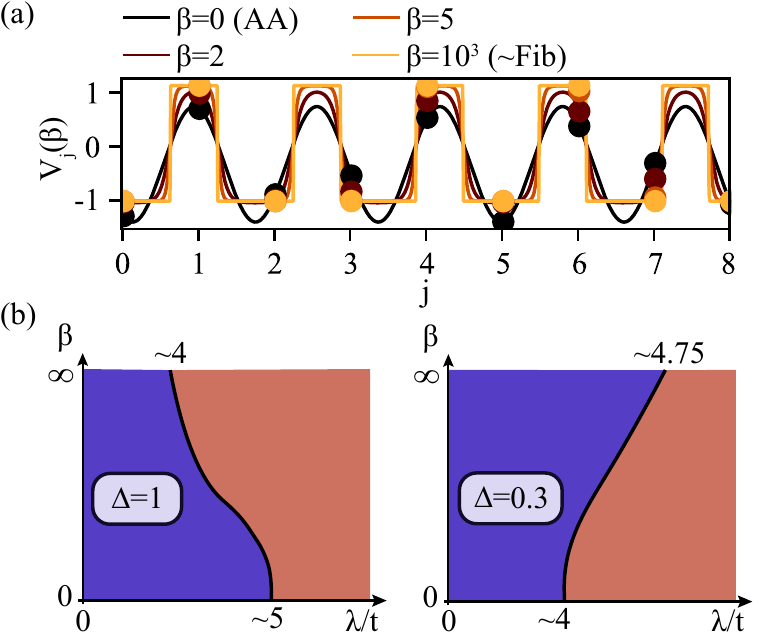}
    \caption{The i-IAAF model and its many-body localization properties.
        (a) Quasiperiodic IAAF potential as a function of space [cf.~Eq.~\eqref{eq:potential}] for different values of $\beta$. The continuous function is plotted and its discrete sampling is marked by dots.
        (b) Sketch of the many-body localization phase diagrams for the i-IAAF [cf.~Eq.~\eqref{eq:TB_hamiltonian}], established in this work for $\Delta=1$ and $\Delta=0.3$. The non-ergodic MBL phase and the delocalized ergodic phase are marked in red and blue, respectively.
    }
        \label{fig:intro}
\end{figure}

The function $V_j(\beta)$ is quasiperiodic [see Fig.~\ref{fig:intro}(a)] with an irrational spatial modulation frequency taken as the inverse of the golden mean, $b=2/(1+\sqrt{5})$. The parameter $\phi \in [0, 2\pi]$ acts as a global translation of the spatially-modulated potential. 
The tunable parameter $\beta$ provides a knob by which we can interpolate between the two known limiting cases: (i) $\beta \rightarrow 0$ yields the Aubry--Andr\'{e} cosine modulation~\cite{aubry1980, harper1955}, up to a constant shift in energy, and (ii) $\beta \rightarrow \infty$ corresponds to a step function switching between $\pm 1$ values according to the Fibonacci sequence~\cite{kohmoto1983, ostlund1983}. 
The interpolating function \eqref{eq:potential} was introduced in Ref.~\cite{kraus2012} for proving the topological equivalence of The AA and Fibonacci models. The advantage of such an interpolation is its simplicity, more precisely, it contains a single tuning parameter which then transforms a cosine function into a steplike function, as shown in Fig.~\ref{fig:intro}(a). One could think, in principle, of other interpolating functions, but the main thing to keep in mind is reaching the correct limits of the AA and Fibonacci model. For other smooth and continuous interpolations, we do not expect the main features of the model to change dramatically. 

Applying the Jordan--Wigner transformation, our model \eqref{eq:XXZ_hamiltonian} is mapped to an interacting one-dimensional model of spinless fermions
\begin{align}
	H = & \sum_{j=1}^{L}  \Bigg[  -t \left( \ccdag[j] \cc[j+1] + \ccdag[j+1] \cc[j] \right)  
	  +  \lambda V_j (\beta) \left( n_j-\frac{1}{2} \right)   \nonumber  \\
	& +  \Delta \left( n_j-\frac{1}{2} \right) \left(n_{j+1}-\frac{1}{2} \right) \Bigg] \, ,
	\label{eq:TB_hamiltonian}
\end{align}  
where $\ccdag[j]$ and $\cc[j]$ are fermionic creation and annihilation operators, $n_j = \ccdag[j] \cc[j]$ is the local density operator, and $t=-J/2$ is the hopping amplitude. The spatially-modulated magnetic field is transformed to a modulated on-site potential, and the term proportional to $\Delta \geq 0$ acts as a nearest-neighbor repulsive interaction. We dub this model the interacting interpolating Aubry--Andr\'e--Fibonacci (i-IAAF) if $\Delta \neq 0$, and IAAF in the noninteracting case.

The IAAF [$\Delta=0$ in Eq.~\eqref{eq:TB_hamiltonian}] was originally employed for the study of topological properties of quasiperiodic chains~\cite{Kraus2012a, kraus2012, Kraus2013, Verbin2013, Kraus2013comm, Kraus2014, Verbin2015, Kraus2016,kellendonk2019bulk,Zilberberg2021}. On the one hand, a topological charge pump arises in the space spanned by the chain's real dimension $j$ and the synthetic dimension spanned by the so-called pump parameter~\cite{Thouless1983,Kraus2012a, Verbin2015, lohse2016thouless, nakajima2016topological} $\phi$. Correspondingly, topological boundary modes cross the bulk gaps as a function of $\phi$~\cite{Kraus2012a,Verbin2015}. Similarly, for modulated interacting chains, boundary modes cross the many-body excitation gaps with $\phi$~\cite{Lado2019}. On the other hand, the quasiperiodic modulation frequency $b$ allows for ``spatial sampling'' of the pump parameter $\phi$ and the 1D chain inherits topological properties from the pump space directly, with related bulk phase transitions and corresponding boundary phenomena~\cite{Kraus2012a, kraus2012, Verbin2013, Kraus2013comm, Kraus2016,kellendonk2019bulk}.

Recently, the IAAF model was shown to exhibit interesting localization properties~\cite{strkalj2020}, which we briefly outline in Section~\ref{sec:nonint_IAAF}.
In this work, we concentrate on the many-body localization (MBL) properties of the i-IAAF \eqref{eq:TB_hamiltonian} as it interpolates, with $\beta$, between the Aubry--Andr\'{e} and Fibonacci limits, see Fig.~\ref{fig:intro}(b). 

\subsection{Noninteracting IAAF model} \label{sec:nonint_IAAF}

\begin{figure}[t!]
	\centering
    \includegraphics[width=\columnwidth]{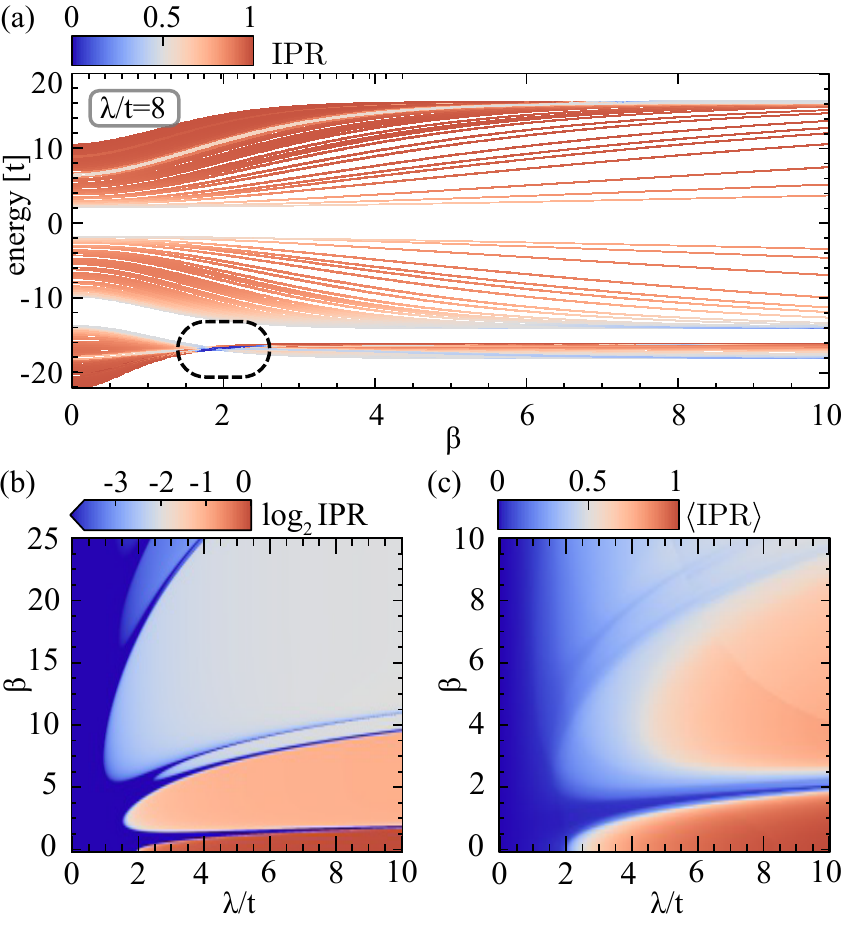}
    \caption{
        The localization properties of the noninteracting IAAF model, cf.~Eq.~\eqref{eq:TB_hamiltonian} with $\Delta=0$. 
        (a)
        The inverse participation ratio (IPR) for every single-particle eigenstate in the spectrum as a function of $\beta$ at constant $\lambda/t=8$. The dashed oval marks the region where the lowest band-bundle of eigenstates becomes extended.
        (b)
        Localization phase diagram of the ground state obtained by the IPR discribed in the main text.
        (c)
        Localization phase diagram obtained by averaging the IPR over the lowest band-bundle. For all plots, we used a system with $L=144$ sites.
    }
        \label{fig:noninteracting}
\end{figure}

Before considering the full interacting model~\eqref{eq:TB_hamiltonian}, we summarize the salient localization properties of its noninteracting version ($\Delta=0$)~\cite{strkalj2020}, see Fig.~\ref{fig:noninteracting}. 
At the $\beta=0$ limit, the noninteracting AA model is known to exhibit a pure spectrum localization transition at~\cite{aubry1980} $\lambda_{\rm C}/t=2$, where all eigenstates of the model change from extended to localized. In the opposite limit, $\beta \rightarrow \infty$, the noninteracting Fibonacci model exhibits critical behavior~\cite{kohmoto1983,Kohmoto1986,Kohmoto1987}, namely the eigenstates decay in space as an inverse power law (critically localized eigenstates) for any value of $\lambda$. 

Along the deformation~\eqref{eq:potential} as a function of $\beta$, the IAAF potential becomes steeper, thus producing a stronger forcing on particles moving in such a potential. As a result, the region, where the eigenstates are purely extended, shrinks. Furthermore, starting from a localized AA phase, where every state is localized on a single site, and upon increasing $\beta$, the potential~\eqref{eq:potential} pushes the states of the lowest energy band to resonance. For a chosen $\lambda/t=8$, the first resonance occurs at $\beta\sim 2$, where the states hybridize and become extended, leading to a mobility edge scenario, which is peculiar for MBL physics~\cite{Potter2014,xu2019,Padavic2021}. Increasing $\beta$ further to $\beta > 2$, the states in the lowest band localize once again. This resonance cascade repeats for higher $\beta$ until all states hybridize to form the critical states of the Fibonacci chain. In Fig.~\ref{fig:noninteracting}(b), we show the cascade for the ground state. Note that the delocalization due to hybridization followed by localization involves a doubling of the localization length, and it does not occur simultaneously for all states in the spectrum.

To visualize the aforementioned localization properties of the IAAF model, we numerically diagonalize a finite chain and evaluate the inverse participation ratio (IPR) for each eigenstate of the model~\eqref{eq:TB_hamiltonian}. The IPR of an eigenstate $\psi(E_n)$ with eigenenergy $E_n$ is given by
$\mathrm{IPR}(E_n) = \sum_{j=1}^{L} |\psi_j(E_n)|^4 / \sum_{j=1}^{L} |\psi_j(E_n)|^2$.
In the regime where the $n$-th eigenstate $\psi(E_n)$ is extended, the IPR is proportional to the inverse of the system length, i.e., $\mathrm{IPR}(E_n)=1/L$ which approaches zero in the thermodynamic limit. On the other hand, if the $n$-th eigenstate is exponentially localized on $N$ sites, its IPR is equal to $1/N$ and remains finite even for an infinite system size $L \rightarrow \infty$.

In Fig.~\ref{fig:noninteracting}(a), we plot the IPR of every eigenstate as a function of $\beta$ for a constant $\lambda/t$, and mark the appearance of delocalized states in the lowest band-bundle of the spectrum. In Fig.~\ref{fig:noninteracting}(c), we show the model's localization phase diagram for the lowest band-bundle of the energy spectrum, marked in Fig.~\ref{fig:noninteracting}(a). We use the averaged inverse participation ratio, $\expval{\mathrm{IPR}} = (1/m) \sum_{n=1}^m \mathrm{IPR}(E_n)$, whose value is equal to $0$ when all eigenstates are extended, and $1$, when all of them are localized on one site.  
The cascade of localization-delocalization transitions is visible, although not as clearly as for each eigenstate separately~\cite{strkalj2020}. The strongest delocalization transition happens around $\beta \sim 2$ for constant $\lambda/t=8$.
Furthermore, the localized region for $2 \lesssim \beta \lesssim 8$ has a large fraction of the states localized on two neighboring sites, which follows from the fact that $\expval{ \mathrm{IPR} } \approx 1/2$.

\subsection{State of the art of MBL in the Aubry-Andr\'{e} and Fibonacci models}
\label{sec:state_of_art}

Before presenting the main results of this work, we discuss relevant MBL results related to the many-body Aubry--Andr\'{e} and Fibonacci models. We also use this opportunity to summarize the computational methods that are often employed to study MBL in disordered or in quasiperiodic models. We first discuss numerical methods and emphasize their advantages and disadvantages. 

We start with exact diagonalization (ED)~\cite{iyer2013}: its main advantage is that it is exact, providing solutions for eigenenergies and eigenstates of interacting Hamiltonians. The exact nature of ED comes at a price:  only very small systems ($\sim 24$ spin-$1/2$ sites) are accessible and the outcome is riddled with finite-size effects, rendering the estimation of the system's behavior in the thermodynamic limit difficult to extract. Using ED, previous works investigated signatures of MBL in the behavior of the one-body density matrix~\cite{bera2017}, statistics of the many-body spectrum~\cite{oganesyan2007}, entanglement entropy~\cite{Bardarson2012,Serbyn2013, Laflorencie2016, abanin2019}, as well as the dynamics of entropy and imbalance after quenching the system~\cite{doggen2018,abanin2019}.

Another class of numerical methods involves a description using matrix product states (MPS), which are variational Ansätze for the many-body wave function. MPS approximate many-body states by truncating the entanglement spectrum, which works especially well in one dimension due to the zero-dimensional boundary for entanglement's area law scaling. There are several MPS-based algorithms for the dynamics of quantum systems \cite{paeckel2019}, such as the time-dependent density matrix renormalization group (tDMRG) \cite{Daley2004,White2004}. Here, we will use  the time-dependent variational principle (TDVP), as recently generalized to matrix product states~\cite{haegeman2011, haegeman2016}. Using this method, one can study large systems with $O(100)$ sites~\cite{doggen2021}, which is crucial for exploring MBL phases. An alternative MPS-based approach~\cite{Znidaric_2010} employs a boundary-driven Lindblad equation. This approach is especially efficient for studying dynamical states close to pure thermal (and hence highly-entangled in the $S^z$ basis) states. Unfortunately, it is therefore less suited for strongly localized and weakly entangled systems. Hence, it was used to study the interacting Fibonacci model only in the limit of relatively weak fields~\cite{varma2019}.

We now outline relevant MBL results for the many-body AA and Fibonacci models, starting with the AA model. In Refs.~\cite{iyer2013, bera2017, lev2017}, ED was used to analyse relatively small chains with $\sim 20$ sites. Initially, an MBL transition was identified for $\Delta=1$ at an estimated critical AA potential strength~\cite{iyer2013} of $2\lesssim \lambda_{\rm C}/t \lesssim 5$. 
In later works, by studying the spectrum of eigenvalues of the one-body density matrix, an MBL transition at $\lambda_{\rm C}/t \approx 4$ was obtained~\cite{bera2017} for $\Delta=1$, while from the statistics of the spectrum of eigenfunctions of the Hamiltonian the transition at $\lambda_{\rm C}/t \gtrsim 3$ was inferred~\cite{lev2017}. 
In Ref.~\cite{doggen2019}, one of us obtained a critical AA potential strength of $\lambda_{\rm C}/t \approx 4.8$ from the dynamics of a quenched imbalance using TDVP. The analysis was performed on long chains with $L=50$ sites.

The interacting Fibonacci model was first explored in Refs.~\cite{vidal1999,vidal2001}. Combining analytical methods, such as bosonization and renormalization group, the low-temperature properties of the model were studied. It was found that the model shows anomalous transport properties with a scaling exponent depending on the interaction strength and the position of the Fermi level.
The high-temperature regime, which we consider in this paper, was recently studied in Refs.~\cite{mace2019, varma2019, Chiaracane2021}.
It was shown~\cite{mace2019} that the interacting Fibonacci model with $\Delta=1$ undergoes an MBL transition at $4 \lesssim \lambda_{\rm C}/t \lesssim 7$. 
Specifically, ED with exact Krylov-space approach was applied to chains of length up to 24 sites, and both static probes, e.g., spectral statistics and scaling of the entanglement entropy, as well as dynamic ones, e.g., entanglement growth and evolution of local observables after a quench, were investigated. 
Further results~\cite{varma2019} using tDMRG, involved boundary-driven Lindblad dynamics, as well as unitary dynamics, suggesting that for $\Delta \lesssim 0.5$, the model is diffusive for all potential strengths, while for $\Delta \gtrsim 0.5$ there exists an interaction-induced subdiffusive regime that persists at least up to $\lambda/t=3$. Recently, by studying the R\'{e}nyi participation entropy and the occupation number at half chain using ED, it was proposed~\cite{Chiaracane2021} that even at a weak interaction ($\Delta \approx 0.5$) an MBL phase should occur within the interval $4<\lambda_{\rm C}/t<8$. 

Many-body interactions bear a somewhat opposite impact on the AA and Fibonacci models.
On one hand, in the AA model, the interactions tend to delocalize the system and shift the localization transition towards higher values of $\lambda/t$ compared with the noninteracting case~\cite{lev2017}. This trend can be qualitatively understood as follows: (i) recall that the localization transition in the noninteracting limit is homogeneous throughout the whole spectrum, and occurs at  $\lambda/t=2$, where the model is self-dual; (ii) the self-duality condition is fragile and is broken by the interactions, even at the mean-field level~\cite{xu2019}; (iii) a mobility edge thus appears, i.e., around $\lambda/t=2$, the spectrum of the weakly interacting AA model contains both extended and localized states; and (iv) the extended states can act as an effective bath to the localized states and delocalize them~\cite{Potter2014,Gopalakrishnan2014}. 

On the other hand, in the Fibonacci model, the interactions destroy critical states and localize the system at large enough $\lambda/t$. 
One possible explanation for such an effect~\cite{mace2019} involves the Fourier components of the Fibonacci potential [$\beta \to \infty$ limit of Eq.~\eqref{eq:potential}], namely, $V(k) = \sum_{j=1}^{L} V_j e^{-i k j b} \sim 1/k$. The  slow decay of the potential is responsible for the model's noninteracting critical eigenstates,  which decay as a power law in real space. 
The interactions on mean-field level introduce new terms in the potential and therefore change the Fourier components to~\cite{mace2019} $V(k) \sim 1/k^{\alpha}$ with $\alpha>1$. Quasiperiodic models with such a Fourier space behavior exhibit localization-delocalization transitions~\cite{Monthus2019}.

%=====================%
\section{Results   \label{sec:results}}
To explore the localization properties of the i-IAAF model~\eqref{eq:TB_hamiltonian}, we examine quench dynamics. As an initial state, we take a N\'{e}el state: $\ket{\psi(t=0)} = \ket{\uparrow, \downarrow, \ldots, \uparrow, \downarrow}$, which is in the $\sum_j \langle S^z_j \rangle \equiv \sum_j \langle \psi| S^z_j |\psi\rangle = 0$ spin sector [Eq.~\eqref{eq:XXZ_hamiltonian}], and corresponds to half-filling in the particle picture [Eq.~\eqref{eq:TB_hamiltonian}]. The N\'{e}el state resembles a mid-band Bloch state that  thermalizes quickly in the absence of disorder~\cite{dalessio2016}. Given the initial state, we employ the time-dependent variational principle (TDVP)~\cite{haegeman2011,haegeman2016} to simulate the time evolution of relatively-long chains ($L=50$ throughout this section).  Our approach is similar to that of recent works~\cite{doggen2018,doggen2019,doggen2020,doggen2021,doggen2021b}. 

The observable that we concentrate on is the spin imbalance defined as
\begin{align}
    \mathcal{I}(\tau) = \frac{1}{L} \sum_{j=1}^L (-1)^j \expval{S_j^z(\tau)}\,,
    \label{eq:imbalance}
\end{align}
where $\tau$ is the simulation time in units of $J^{-1}$.
It quantifies the local memory of the N\'{e}el initial state; the value of the imbalance at the initial time is $\mathcal{I}(\tau=0)=1$.
For a system in a localized phase, the imbalance saturates to a constant value at long times.
If the system is in the ergodic phase, the imbalance vanishes in the long-time and thermodynamic limit.
In the case of random on-site disorder (corresponding to Anderson localization in the noninteracting limit), the imbalance averaged over many disorder realizations decays as an inverse power law close to the MBL transition point~\cite{luitz2017}
\begin{align}
    \overline{\mathcal{I}(\tau)} \propto \tau^{-\gamma} \, ,
    \label{eq:imbalance_decay}
\end{align}
where the bar denotes disorder averaging and the exponent $\gamma$ depends on the strength of the disorder. Deep inside the MBL phase, memory of the initial state persists at all times and therefore $\gamma\rightarrow 0$. 

In the quasiperiodic case, due to the absence of Griffiths effects, which depend on the appearance of rare regions, the decay does not exactly obey a power-law form with subdiffusive exponents at late times.
Nonetheless, dynamics in the localized regime will still saturate to a finite imbalance, resulting in a vanishing $\gamma$. Note that the counterpart of disorder averaging in the quasiperiodic case is averaging over the phases $\phi$.

This section is organized as follows: in Section~\ref{subsection:Fibonacci}, we report our results for the interacting Fibonacci model [the $\beta \rightarrow \infty$ limit of Eq.~\eqref{eq:TB_hamiltonian}] and compare them with contemporary literature on this limit~\cite{mace2019,varma2019,Chiaracane2021}. In Section~\ref{subsection:i-IAAF}, we obtain the phase diagram of the i-IAAF model with $\Delta=1$ and $\Delta=0.3$, and in Section~\ref{subsection:anomalous}, we concentrate on a specific region in the phase diagram, where anomalous localization-delocalization behavior occurs by tuning the interaction strength from $\Delta=0$ to $\Delta=1$.  

\subsection{Fibonacci model   \label{subsection:Fibonacci}}
\begin{figure}[t!]
	\centering
    \includegraphics[width=\columnwidth]{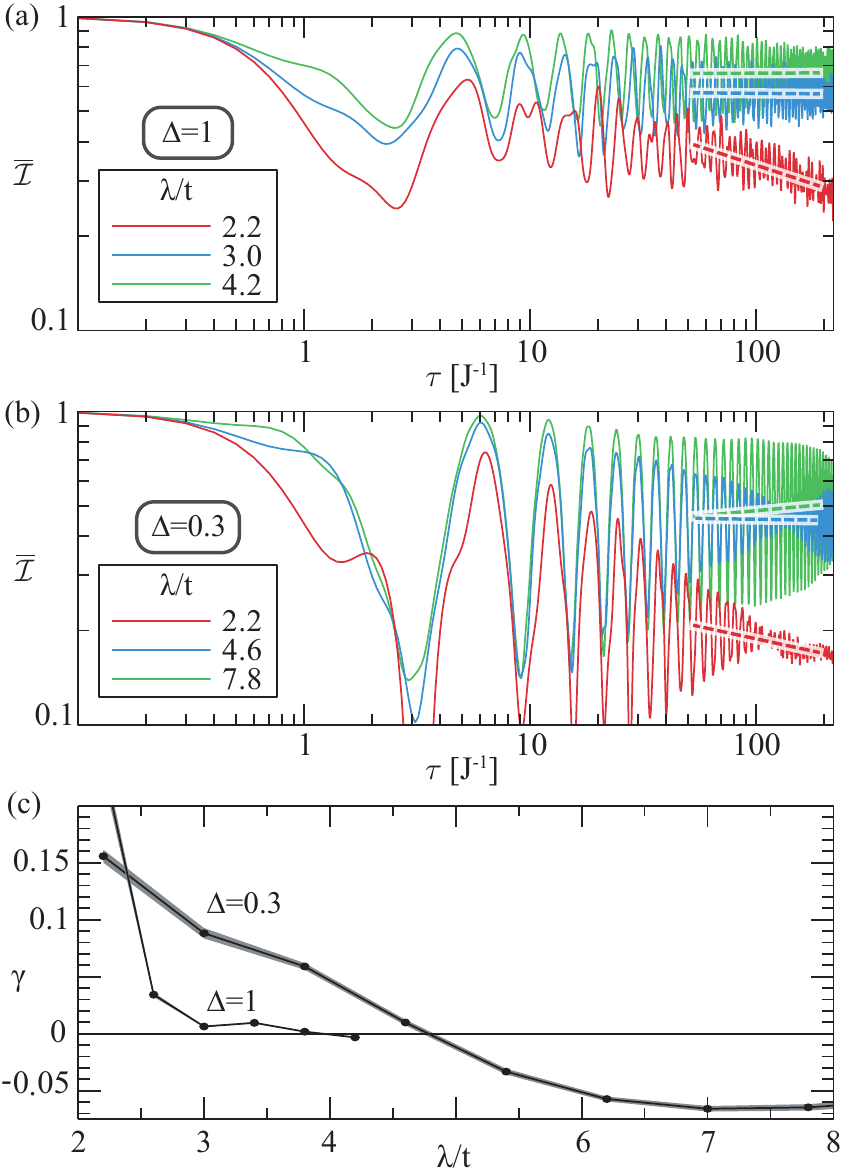}
        \caption{ 
        Numerical simulation of the MBL transition in the Fibonacci model.
        We show the time evolution of the averaged imbalance~\eqref{eq:imbalance_decay} using TDVP for three amplitudes, $\lambda/t$, of the quasiperiodic potential and two interaction strengths (a) $\Delta=1$ and (b) $\Delta=0.3$. In both cases, for $\lambda/t=2.2$ the averaged imbalance decays (red line), indicating that this point lies in the ergodic phase. For the other two presented plots (blue and green lines), $\overline{\mathcal{I}}$ saturates over time to a constant. 
        (c) The inverse power law coefficient~\eqref{eq:imbalance_decay} fitted to $\overline{\mathcal{I}(\tau)}$ over a finite fitting time window $\tau\in[50,200]$ [dashed lines in (a) and (b)] as a function of $\lambda/t$. 
        For all plots we use $L=50$ sites and averaged the imbalance over 36 realizations with different $\phi$ taken from the interval $\phi \in [0, 2\pi]$. 
        The bond-dimension used for all plots is $\chi=64$ and the time step is $\delta \tau=0.1$.
        Error bars are $1\sigma$ intervals based on a bootstrapping procedure~\cite{Efron1979}.
        }
        \label{fig:fibonacci}
\end{figure}

We start by revisiting the interacting Fibonacci model, i.e., the $\beta \rightarrow \infty$ limit of Eq.~\eqref{eq:TB_hamiltonian}. We calculate the time evolution of the imbalance~\eqref{eq:imbalance} for several values of the potential strength $\lambda/t$, and show the result in Fig.~\ref{fig:fibonacci}. 

\textit{Strong interactions} -- The  $\Delta=1$ case is shown in Figs.~\ref{fig:fibonacci}(a) and (c).
In Fig.~\ref{fig:fibonacci}(a), we observe that the imbalance exhibits strong oscillations as it evolves in time, showing revivals associated with short-scale oscillations of the particles within their close environment. At the same time, a clear distinction manifests in the long-time behavior, where, for $\lambda/t \approx 2.2$, the imbalance decays roughly as an inverse power law, while for $\lambda/t \geq 3$ it saturates to a constant value. We fit the long-time trend with an inverse power law behavior [cf.~Eq.~\eqref{eq:imbalance_decay}], and obtain the exponent $\gamma$, see dashed lines in Fig.~\ref{fig:fibonacci}(a). Repeating this procedure for several values of $\lambda/t$, we plot the fitted $\gamma$ in Fig.~\ref{fig:fibonacci}(c). 
Initially, the exponent decreases with increasing $\lambda/t$. This decrease saturates at $\lambda/t \approx 3$, while at $\lambda/t \approx 4$ the imbalance does not appear to decay on the computationally-available evolution timescales. Hence, we estimate that an MBL transition occurs within the interval $3 \leq \lambda_{\rm C}/t < 4$. As a further confirmation of our finding, in Appendix~\ref{appendix_A}, we show similar results for a numerically exact quench of a shorter chain with $L=16$ sites and bond dimension $\chi=256$.

\textit{Weak interactions} -- We now turn to weaker interactions, in order to compare with the results presented in Ref.~\cite{varma2019}. It was suggested that the Fibonacci model is diffusive for $\lambda/t \leq 3$ and at low interaction strengths, $\Delta<0.5$, while at $\Delta=0.5$ a subdiffusive phase appears, but no MBL phase was predicted. In Figs.~\ref{fig:fibonacci}(b) and (c), we report the outcome of our numerical quench for a low-interaction strength $\Delta=0.3$. Here, the imbalance decays faster and exhibits slower oscillations relative to the former $\Delta=1$ case. Moreover, the appearance of an MBL phase occurs only for much higher $\lambda/t$ values, where even for $\lambda/t=4.6$, we observe a slow decay. In Fig.~\ref{fig:fibonacci}(c), we show the fitted power-law exponent as a function of the potential strength. A clear memory of the initial state remains for $\lambda_{\rm C}/t \gtrsim 4.75$. Comparison with exact numerics for a chain of $L=16$ sites and $\chi=256$ (see Appendix~\ref{appendix_B}) yields a similar value of $\lambda_{\rm C}/t \gtrsim 4.5$. For even weaker interactions, i.e., $\Delta=0.1$, we do not observe signatures of an MBL phase transition for potential strengths $\lambda/t \leq 10$. 

\textit{Source of oscillations} -- We now comment on the oscillations present in Figs.~\ref{fig:fibonacci}(a) and (b). Note that, unlike true disorder or the AA case, the amplitude of the oscillations does not reduce with averaging over a large number of different quasiperiodic potential realizations. The reason is that the Fibonacci potential has only $L/2$ possible configurations~\cite{mace2019}, where $L$ is the number of sites in the system. A possible way to reduce the amplitude of the oscillations is to use random product states as the initial state instead of the N\'{e}el state used in this work, which goes beyond the scope of this work. Instead, we explore here the behavior of these oscillations stemming from the initial N\'{e}el state. Note that the persistent oscillations and the memory of the initial state can lead to negative values of $\gamma$, similar to what is observed in the AA model \cite{doggen2019}.

The oscillations are irregular for a weak potential, i.e., in the ergodic phase, while for strong quasiperiodic potentials they become more regular with an approximately constant period. Such pronounced oscillations are due to the shape of the potential, which contains many occurrences of nearest-neighbor pairs with  degenerate on-site energies. In the initial Ne\'{e}l state, there is only one particle populating any such pair of states. That particle will dominantly hop between the nearest neighbor degenerate states and produce oscillations in the imbalance. This simplified picture is valid for strong potentials and weak interaction, and breaks in the opposite limit of weak potentials and strong interaction. 

The periods of oscillations extracted from the averaged imbalance in the localized phase [see Fig.~\ref{fig:fibonacci}(a) and (b)] are $T_{\Delta=1} \approx 2 \pi / \sqrt{2} J^{-1}$ and $T_{\Delta=0.3} \approx 2\pi J^{-1}$ for  $\Delta=1$ and $\Delta=0.3$ interaction strengths, respectively. From the periods it is possible to extract an effective hopping $\tilde{t} \propto 1/T$, see Ref.~\cite{barmettler2010} for a full derivation of the imbalance in the noninteracting and clean case. The ratio of effective hopping from Figs.~\ref{fig:fibonacci}(a) and (b) is then $\tilde{t}_{\Delta=1}/\tilde{t}_{\Delta=0.3} \approx \sqrt{2}$.
To explain the difference between two effective hoppings, in Appendix~\ref{appendix_C} we present a simple model that describes how interactions change the particles dynamics within the aforementioned degenerate nearest-neighbor pairs. This simple analytical model yields $\tilde{t} = t \, \sqrt{1+(\Delta/J)^2}$, which agrees well with the numerical results from Figs.~\ref{fig:fibonacci}(a) and (b). For the case of weak interactions [Fig.~\ref{fig:fibonacci}(b)], the effective hopping is given by $\tilde{t}=t$, i.e., the oscillations appear to stem from the single-particle dynamics, while in the case of strong interactions [Fig.~\ref{fig:fibonacci}(a)], the effective hopping is renormalized to $\tilde{t} \approx \sqrt{2} t$, suggesting that interactions increase the effective kinetic energy of the particles. To conclude this discussion, the oscillations in the MBL phase indeed stem from an ensemble of effective particles hoppings between two degenerate neighboring sites. 

\subsection{Interacting IAAF model   \label{subsection:i-IAAF}}
\begin{figure}[t!]
	\centering
    \includegraphics[width=\columnwidth]{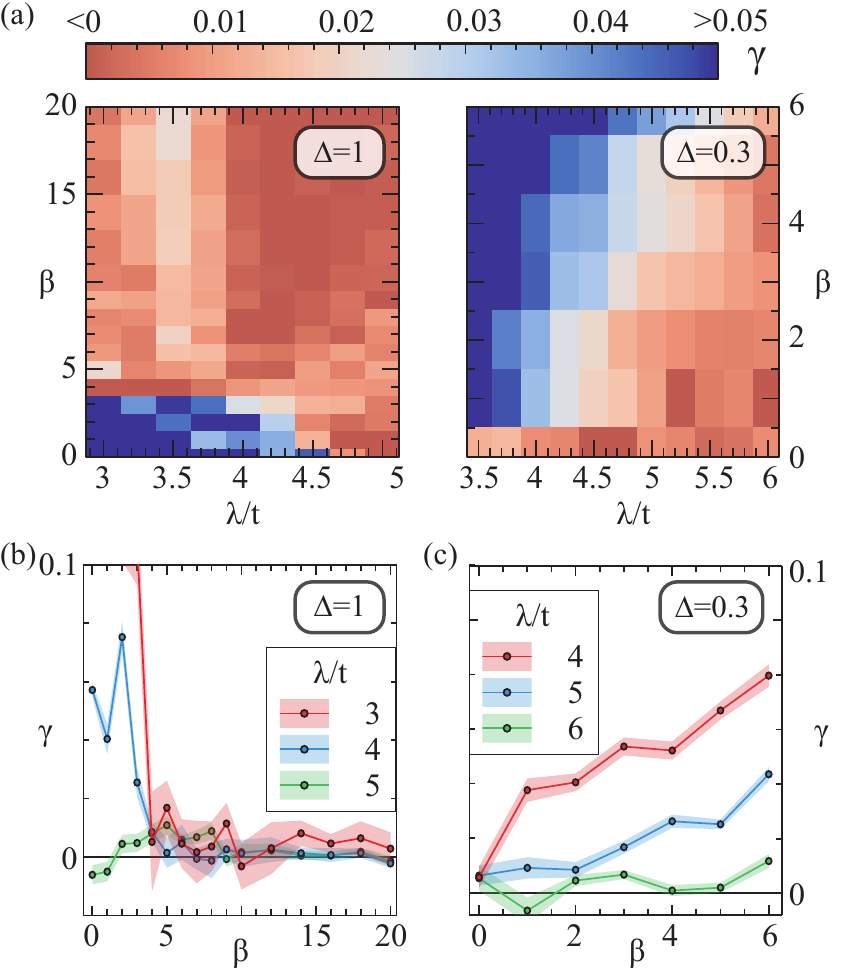}
        \caption { 
        Numerical simulation of the MBL transition in the i-IAAF model.
        (a)
        Many-body localization phase diagram obtained from the exponent $\gamma$ [cf.~Eq.~\eqref{eq:imbalance_decay}], for (left) strong interaction, $\Delta=1$, and (right) weak interaction, $\Delta=0.3$.
        The bottom panels show $\gamma$ as a function of beta for selected values of $\lambda/t$ for (b) $\Delta=1$, and (c) $\Delta=0.3$. 
        The numerical specifications used are the same as in Fig.~\ref{fig:fibonacci}, with the distinction that the fitting time window now depends on $\lambda/t$ as described in Appendix~\ref{appendix_B}.
        }
    \label{fig:MB_phase_diagram}
\end{figure}

We now address the i-IAAF model with strong and weak interaction, $\Delta=1$ and $\Delta=0.3$, respectively. For the strong interaction, the MBL transition in AA and Fibonacci models occurs at different potential strengths $\lambda_{\rm C}/t \approx 5$ and $\lambda_{\rm C}/t \approx 4$, respectively. Note that in this case, the Fibonacci MBL transition occurs for a lower $\lambda_{\rm C}/t$ than that of the AA. Interestingly, for the weak interaction case, the order inverts and we find in the AA case $\lambda_{\rm C}/t \approx 4$ while for the Fibonacci $\lambda_{\rm C}/t \approx 4.75$. 
It is, therefore, our goal to explore how the two transition points are connected once $\beta$ is tuned from AA ($\beta=0$) to the Fibonacci model ($\beta \rightarrow \infty$), for both weak and strong interactions, cf.~Fig.~\ref{fig:intro}(b). To answer these questions, we again analyze the dynamics of the averaged imbalance, see Eq.~\eqref{eq:imbalance_decay} and Fig.~\ref{fig:MB_phase_diagram}(a) for the fitted exponent $\gamma$ as a function of $\lambda/t$ and $\beta$. 

For $\Delta=1$, starting from the MBL phase of the AA model $\lambda/t \geq 5$ and increasing $\beta$, $\gamma$ always remains around 0 and no transition to an ergodic phase is observed. 
On the other hand, starting from the ergodic phase of the AA, an MBL transition takes place for higher values of $\beta$. 
The value of $\lambda_{\rm C}/t$, where the transition occurs, decreases as a function of $\beta$. Such a behavior is reminiscent of the noninteracting case, where the impact of the potential becomes more pronounced with $\beta$, and the region of extended states shrinks, see Figs.~\ref{fig:noninteracting}(b) and (c). Note, however, a crucial difference between the noninteracting and interacting cases is the absence of localization-delocalization transitions with changing $\beta$ above the critical localization point of the AA model.

For $\Delta=0.3$, at $\beta=0$ and $\lambda/t \gtrsim 5$, the situation is similar to the 
strong-interaction case: the exponent always remains around zero and no transition to an ergodic phase is observed, see Fig.~\ref{fig:MB_phase_diagram}(a). On the other hand, starting from the localized phase of the AA model at $4 \leq \lambda/t < 5$, a transition to the ergodic phase is observed at the value of $\lambda/t$ that depends on $\beta$; the transition point moves to higher values of $\lambda/t$ with increasing $\beta$. Note that we explore only up to $\beta=6$ in the phase diagram for weak interaction. For $\beta \geq 6$, the decay of the imbalance with time does not change, but the long-period oscillations make it difficult to properly determine the exponent, cf.~Fig.~\ref{fig:fibonacci}(b), where  similar fitting challenges appear in this limit. For example, the imbalance for both $\lambda/t=4.6$ and $\lambda/t=7.8$ in Fig.~\ref{fig:fibonacci}(b) saturates with additional short-period oscillations, but the value of the fitted exponent is not the same, see Fig.~\ref{fig:fibonacci}(c). 

Despite these technical challenges, we clearly observe that the interaction shifts the critical point toward higher values of $\lambda/t$ in the AA limit, and to lower values in the Fibonacci limit. 
Furthermore, interpolating between the two models (i.e., for finite $\beta$), the critical points at the two limits (AA and Fibonacci) are connected with a continuous line separating the ergodic from the localized phase. As a result, we conclude that MBL transitions will occur also with changing $\Delta$ in the intermediate regions of the interpolation.

Similarly to the Fibonacci case in the previous section, we comment briefly on the oscillatory behavior of the imbalance. While the oscillations are pronounced even at large times in the Fibonacci limit [see Figs.~\ref{fig:fibonacci}(a) and (b)], for finite $\beta$, their amplitude decays with time, see Fig.~\ref{fig:changing_interaction}(b). The reason lies in the shape of the potential. As previously discussed, in the Fibonacci limit, the potential~\eqref{eq:potential} allows for the appearance of pairs of neighboring sites that have the same on-site value [cf.~Fig.~\ref{fig:intro}(a)]. Specifically, the Rabi oscillations of particle hopping between two such sites, which appear many times throughout the chain, give rise to coherent oscillations in the imbalance that persist even at long times. In comparison, for finite $\beta$, the degeneracies of the potential characteristic of the Fibonacci limit are lifted, and, hence, particles hopping between the pairs of sites will exhibit different Rabi oscillations, which results in a reduced amplitude of oscillations in the overall imbalance. Nonetheless, there are still oscillations that are longer-lived than in the case of purely random disorder, since certain values of the potential differences between neighboring sites are more likely than others \cite{doggen2019}.

\subsection{Anomalous localization by interaction   \label{subsection:anomalous}}
\begin{figure}[t!]
	\centering
    \includegraphics[width=\columnwidth]{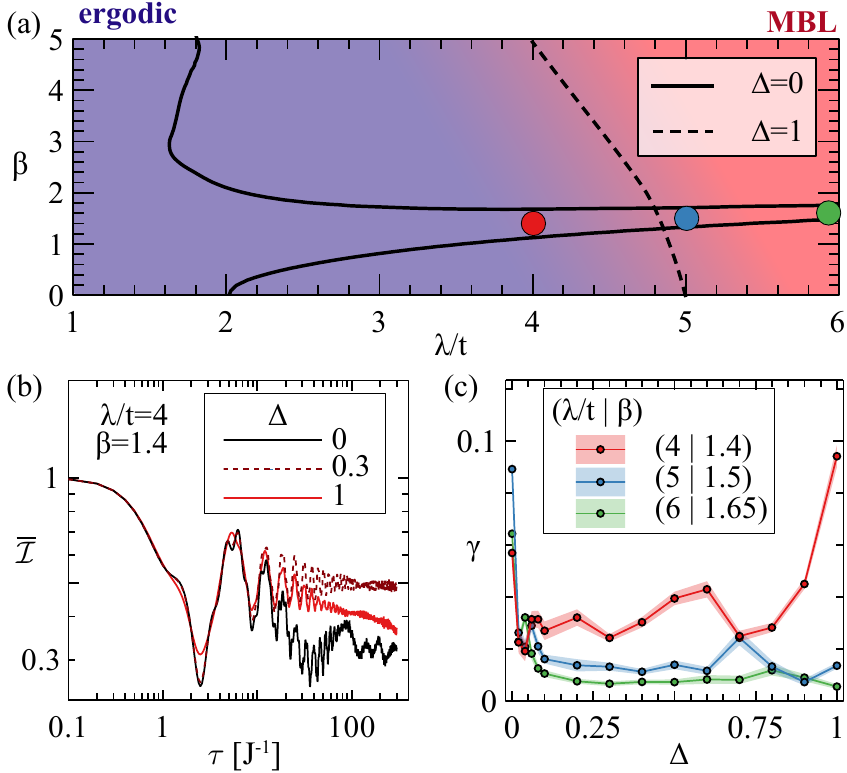}
    \caption { Interaction-dependent localization-delocalization transition.
		(a) 
		Sketch of the overlapped interacting ($\Delta=1$) and noninteracting ($\Delta=0$) localization phase diagrams, cf.~Figs.~\ref{fig:noninteracting} and~\ref{fig:MB_phase_diagram}. 
		Solid lines denote the transition from extended to localized phase in the noninteracting limit, $\Delta=0$, see also Fig.~\ref{fig:noninteracting}(c). Dashed line denotes the MBL transition for $\Delta=1$.
		(b)
		Time evolution of the averaged imbalance~\eqref{eq:imbalance_decay} at a constant $\lambda/t=4$ and $\beta=1.4$ [red point in (a)]. 
		(c) The exponent of the averaged imbalance decay~\eqref{eq:imbalance_decay}, as a function of interaction strength $\Delta$ for several points in the phase diagram [also marked in (a)]. We pick points that lie within the delocalized-phase sliver of the noninteracting model, i.e., situated at finite $\beta\sim 2$ and at $\lambda/t>2$.
		System size in all plots is $L=50$, the number of different realizations of the quasiperiodic potential is 36, and the inverse power law is fitted in a time window [10, 200].
		The bond dimension used for all plots is $\chi=64$ and the time step is $\delta \tau=0.1$.
		Error bars are $1\sigma$ intervals based on a bootstrapping procedure.}
    \label{fig:changing_interaction}
\end{figure}

After analyzing the overall phase diagrams of the weakly and strongly interacting IAAF model, we now concentrate on the noninteracting delocalized-phase sliver at $\lambda/t>2$ and $\beta\sim 2$, and investigate its evolution when the interaction strength is increased from $\Delta=0$ to $\Delta=1$. We pick a point from the aforementioned region with $(\lambda/t,\beta)=(4, 1.4)$, see the red dot in Fig.~\ref{fig:changing_interaction}(a). This point lies in the ergodic phase of the strongly interacting ($\Delta=1$) model, cf.~Fig.~\ref{fig:MB_phase_diagram}(a).
Fixing the parameters $\lambda/t$ and $\beta$, we calculate the time evolution of the averaged imbalance for three different interaction strengths, $\Delta=0$, $\Delta=0.3$, and $\Delta=1$, see Fig.~\ref{fig:changing_interaction}(b). 

In the noninteracting and strongly interacting regimes, the imbalance decays quickly. Interestingly, for weak interactions, the decay is noticeably slower. In Fig.~\ref{fig:changing_interaction}(c), we plot the exponent~\eqref{eq:imbalance_decay}) as a function of interaction strength, and find that it indeed does not change  monotonously with $\Delta$. Specifically, by increasing the interaction strength, the system first localizes for a weak interaction and then delocalizes at a stronger interaction, which is indicated by a noticeably larger $\gamma$ for $\Delta=0$ and $\Delta=1$, compared to intermediate values of $\Delta$. 

We repeat this procedure for two additional points from the  noninteracting delocalized sliver, see the blue and green points in Fig.~\ref{fig:changing_interaction}(a) and 
the result in Fig.~\ref{fig:changing_interaction}(c). As expected, for $\Delta=0$, the exponent $\gamma$ is large, reflecting the delocalized nature of the noninteracting system, see Fig.~\ref{fig:noninteracting}(a). 
At finite interaction strengths, the exponent reduces in value and after a  critical interaction strength $\Delta_{\rm C}$ remains constant. This indicates that the system's dynamics slows down with increasing  interaction strengths, until the system becomes localized.

%=====================%
\section{Discussion     \label{sec:discussion}}
Let us now collect all the observations from Section~\ref{sec:results} and discuss them from a broader perspective, focusing on qualitative mechanisms that are responsible for the localization properties of the i-IAAF model. 

In the two limiting cases of AA and Fibonacci models, the localization properties have been  discussed in the literature in great detail for both single-particle and many-body cases. The crucial difference between the AA and Fibonacci models is the shape of the potential modulations: in the former, the onsite energies are distributed in the interval $[-\lambda, +\lambda]$, whereas in the latter, they are highly degenerate with only two values $\pm \lambda$. This leads to a localization transition in the noninteracting AA model and a critical spectrum in the Fibonacci model. Since the single-particle properties of the two models are very dissimilar, many-body interactions introduce distinct effects in them.

As already mentioned in Section~\ref{sec:state_of_art}, in the AA model, interactions shift the localization phase transition towards higher values of $\lambda/t$ due to dephasing, similar to the interacting Anderson model with random disorder. On the other hand, in the Fibonacci limit, the interactions destroy the fragile criticality caused by the numerous degenerate on-site terms, and introduce a localization transition. This can be understood qualitatively on a mean-field level, where interactions introduce new onsite terms, i.e., shift the onsite energies, and break the degeneracy of the potential. 
Once the degeneracy is lifted, the hopping of particles is suppressed, and the system becomes localized for large enough $\lambda/t$.
It is, therefore, expected that by increasing the interaction strength $\Delta$, the system will localize more easily, i.e., for lower values of $\lambda/t$. Such behavior is observed in Fig.~\ref{fig:fibonacci}. 

Now, we turn to the i-IAAF model. In its noninteracting limit, the model shows rich localization properties with a cascade of localization-delocalization transitions, as shown in Fig.~\ref{fig:noninteracting}. Let us concentrate on the first delocalization transition at finite $\beta$. Starting from a localized phase in the AA limit, the deformation of the potential from a cosine to a step-like function with increasing $\beta$ brings the low-energy states into resonance. 
Once in resonance, the states hybridize for any finite hopping and delocalize.
As shown in Fig.~\ref{fig:noninteracting}(a), all states in the lowest band-bundle are delocalized, and the singe-particle spectrum has a nontrivial mobility edge. The nontrivial mobility edges repeat at different positions in the spectrum for higher $\beta$.
It is important to note that the width of the delocalized band-bundle shown in Fig.~\ref{fig:noninteracting}(a) reduces with increasing $\lambda/t$, since the hopping between the sites is effectively reduced. As a consequence, the delocalized regions (slivers) in the phase diagram shrink at higher $\lambda/t$. 

Adding interactions alters the localization properties of the model as shown in Figs.~\ref{fig:MB_phase_diagram} and~\ref{fig:changing_interaction}. 
There are two main features of the presented many-body phase diagrams: (i) the slope of the MBL phase boundary changes sign with the strength of many-body interactions, and (ii) the delocalized slivers of the noninteracting model disappear once interactions are introduced. 
Point (i) is expected since interactions have opposite effects in AA and Fibonacci limits, as discussed above.
On the other hand, point (ii) is surprising since one would naively expect that the existence of delocalized states in the spectrum of the noninteracting limit will act as a bath to localized states once interactions are turned on, and will eventually delocalize the whole system~\cite{Potter2014,Gopalakrishnan2014}. However, as already mentioned, such delocalized states at finite $\beta$ are the consequence of carefully tuned degeneracy between onsite energies imposed by the deformation of the potential \eqref{eq:potential}, see Figs.~\ref{fig:intro}(a) and~\ref{fig:noninteracting}(a). Similar to the Fibonacci case, interactions can easily lift the degeneracy and localize these slivers. 

Indeed, we recall that the width of the delocalized band-bundle reduces with increasing $\lambda/t$, and, therefore, at higher $\lambda/t$, it should feel stronger effects of interactions, i.e., the delocalized states should localize for a smaller interaction strength $\Delta$.
Such an effect is observed in Figs.~\ref{fig:changing_interaction}(b) and (c), namely, at the points with higher $\lambda/t$ the decaying coefficient is smaller for all values of interaction strength, indicating stronger localizing effect of interactions for higher $\lambda/t$. 
Furthermore, at some points of the phase diagram that lie inside the delocalized sliver, we find an anomalous behavior with tuning of the interaction strength, where weak interactions tend to localize the system, while stronger interactions delocalize it again, see $(\lambda/t, \beta) = (4, 1.4)$ point in Fig.~\ref{fig:changing_interaction}. The localization at weak interactions can be explained with the degeneracy-lifting  argument above, while the delocalization could follow from the same dephasing mechanism that is present in the AA limit. Further studies are needed to better explain and describe this peculiar phenomenon.

%=====================%
\section{Conclusion}
\label{Sec:Conclusion}
To conclude, we have investigated the many-body version of the IAAF model, focusing on the half-filling sector of the model and numerically studying the time evolution of the spin imbalance using the TDVP method~\cite{haegeman2011,haegeman2016}. 
Our main criterion for localization is the saturation of the averaged imbalance at long times. In the ergodic phase, the imbalance decays over time. 
We have observed that for quasiperiodic models with weak interaction, the high degeneracy of the on-site potentials leads to strong oscillations with time in the averaged imbalance. It is, therefore, more difficult to locate a precise MBL transition point in such models. Nevertheless, we have constrained the transition regime, and thus explored the localization-delocalization phase diagram of the interacting IAAF model.

Our work contains three main results. First, we have shown that for strong interaction, $\Delta=1$, the MBL phase transition in the Fibonacci model ($\beta \rightarrow \infty$ limit of the IAAF model) is located at $3 \leq \lambda/t < 4$, which is lower than previously reported values~\cite{mace2019}. Furthermore, we have demonstrated that the MBL transition can survive even for small interactions of $\Delta = 0.3$, and thus complemented the results of Ref.~\cite{varma2019}.
Secondly, we have presented the many-body phase diagram for the IAAF model, which exhibits a monotonous evolution of the MBL phase when going from the AA ($\beta=0$) to the Fibonacci ($\beta=\infty$) limit. In other words, the cascade of delocalization transitions in a single-particle case~\cite{strkalj2020} is destroyed for both weak and strong interactions. 
Lastly, we have observed an anomalous behavior when increasing the interaction strength from $\Delta=0$ to $\Delta=1$ in the parameter sliver where part of the spectrum is delocalized in the noninteracting limit. In other words, in the sliver both extended and localized states exist in the system, leading to a situation where weak interaction tends to localize the system, while stronger interaction drives it to the ergodic phase.

Future work will focus on  energy densities away from half-filling. By studying the model only at the half-filling, we were not able to conclude if the localization properties are homogeneous throughout the spectrum, or whether an interesting cascade structure appears, as in the single-particle case. Moreover, the present study was performed at an infinite temperature, and the behavior with finite temperatures remains unknown. Therefore, one promising direction would be to analytically investigate the i-IAAF model at finite temperatures and for different chemical potential using, e.g., bosonization and renormalization-group approaches~\cite{vidal1999, vidal2001}.

%=====================%
\section*{Acknowledgments}
We thank M.~S.~Ferguson, I.~Ki\v{c}i\'{c}, J.~L.~Lado, A.~D.~Mirlin and D.~G.~Polyakov for fruitful discussions. We acknowledge financial support from the Swiss National Science Foundation, the DFG (project No. GO 1405/6-1), and the RFBR (Grant No. 18-02-01016). 
Numerical calculations in this work have been performed using the TeNPy library \texttt{v0.4.1}~\cite{tenpy}. 

%=====================%
\appendix

\section{Exact numerics for a short chain in the Fibonacci limit \label{appendix_A}}
\begin{figure}[ht]
	\centering
    \includegraphics[width=\columnwidth]{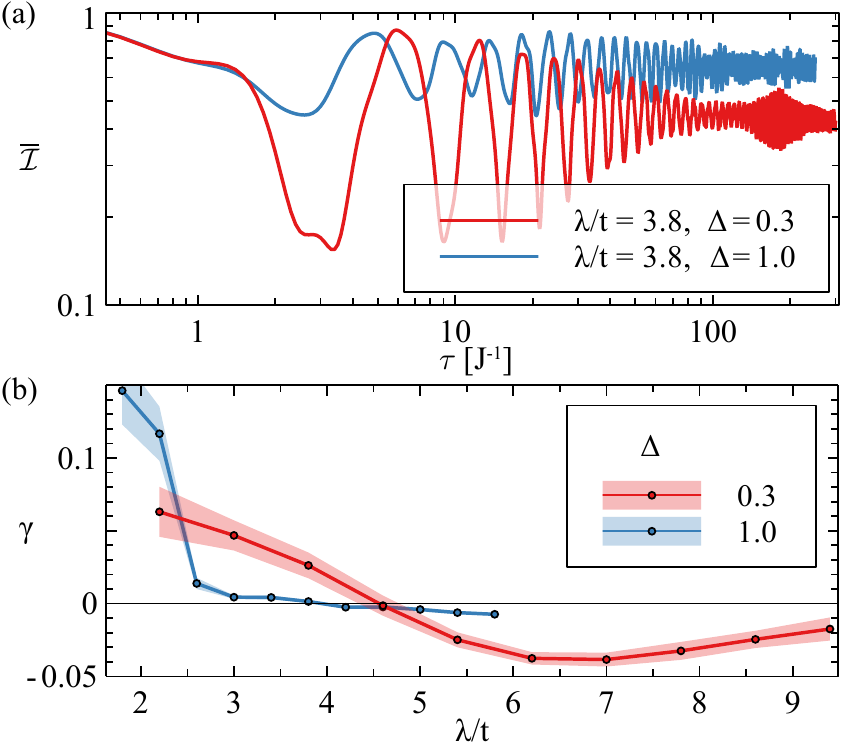}
        \caption { 
        (a) Time evolution of the imbalance~\eqref{eq:imbalance_decay} for two different values of interaction at the constant $\lambda/t=3.8$ in the Fibonacci limit, cf.~Fig.~\ref{fig:fibonacci}.  
        (b) Fitted power-law exponent as a function of the potential strength for the strong and weak interaction cases. 
        We use the following parameters: $L=16$, $\chi=256$, $\delta \tau=0.1$. Error bars are obtained from $1\sigma$ intervals based on a bootstrapping procedure.
        }
        \label{fig:appendix_fibonacci}
\end{figure}

To further confirm the results from Fig.~\ref{fig:fibonacci} from the main text, in Fig.~\ref{fig:appendix_fibonacci}, we show the averaged imbalance and the fitted exponent $\gamma$ for a short chain of $L=16$ sites and large bond dimension $\chi=256$. In this case, the calculation is exact, since, for a system of $L$ sites, non-truncated MPS have a maximum bond dimension of $\chi=2^{L/2}=256$. The behavior of the exponent in Fig.~\ref{fig:appendix_fibonacci}(b) is essentially identical to the one presented in Fig.~\ref{fig:fibonacci}, where we used larger chains with truncation of the bond dimension.  

\section{Convergence of the numerical results   \label{appendix_B}}
In this Appendix, we verify the convergence of our numerical results, namely the convergence of the obtained imbalance decay exponent, which we use to map out the phase diagrams in Fig.~\ref{fig:MB_phase_diagram}. The dynamics of the imbalance is calculated using the TDVP method, whose long-time precision depends on the chosen bond dimension $\chi$. Since the truncation error grows with time, the bond dimension determines the maximal time below which the calculated imbalance does not differ much from its true value. Larger bond dimensions are required in the ergodic phase compared to the MBL phase to reach the same maximum time, see Refs.~[\onlinecite{haegeman2011}, \onlinecite{haegeman2016}, \onlinecite{doggen2018}, \onlinecite{doggen2019}, \onlinecite{doggen2020}] for more details. 

In our case, we use the same bond dimension for the whole phase diagram [see Fig.~\ref{fig:MB_phase_diagram}(a)], implying that we need to change the upper boundary of the fitting time interval as a function of the parameters in the phase diagram to avoid large numerical errors. 
To do so, we choose a few different $\lambda/t$ cuts in phase diagrams shown in Fig.~\ref{fig:MB_phase_diagram}(a), and calculate the exponent for larger bond dimension of $\chi=128$. We assert that the result has converged when for a given fitting time-window, the exponents $\gamma$ in the $\chi=64$ and $\chi=128$ cases do not differ more than the error bars from one another.

\begin{figure}[!t]
	\centering
    \includegraphics[width=\columnwidth]{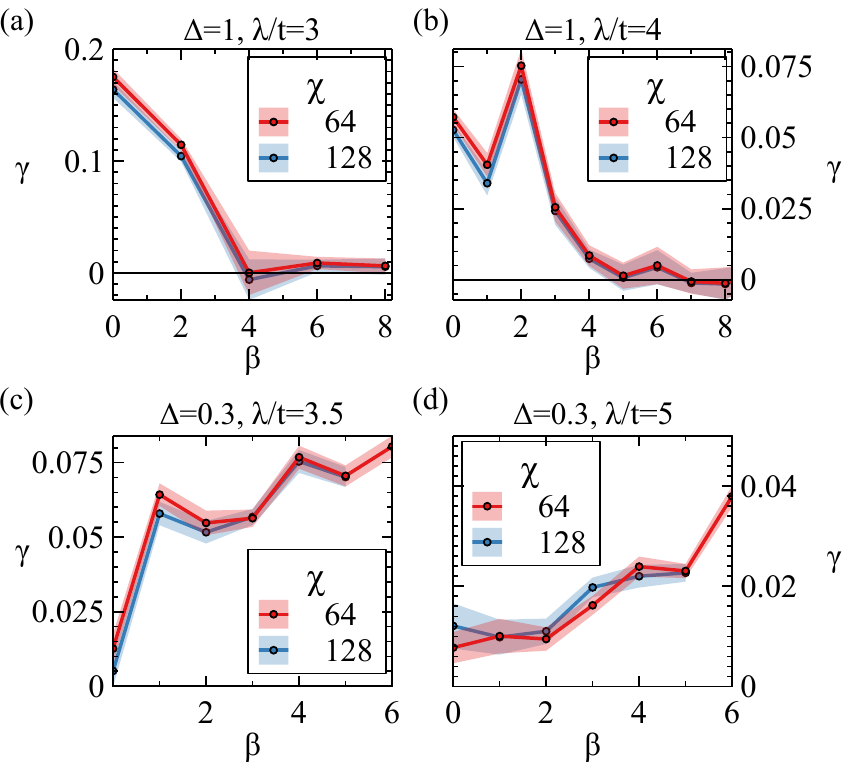}
    \caption{ 
    The power-law exponent $\gamma$ as a function of $\beta$ for two values of a bond dimension $\chi$ and at several values of $\lambda/t$. Panels (a) and (b) show the results for strong interaction, $\Delta=1$, at $\lambda/t=3$ and $\lambda/t=4$, respectively. Dependencies in panels (c) and (d) are calculated for a weak interaction, $\Delta=0.3$, and at $\lambda/t=3.5$ and $\lambda/t=5$, respectively.
    The length of the system is $L=50$ sites, and the time step used for all plots is $\delta \tau=0.1$.
    }
    \label{fig:appendix_comparison}
\end{figure}

The comparison of our results for the two different bond dimensions is presented in Fig.~\ref{fig:appendix_comparison}. For the strong interaction, results with different $\chi$ converge for a fitting time-window $[50, 120]$ used in Fig.~\ref{fig:appendix_comparison}(a), and $[50,200]$ used in Fig.~\ref{fig:appendix_comparison}(b). In the case of weak interaction, the convergence is reached for a fitting time-window $[50, 210]$ in Fig.~\ref{fig:appendix_comparison}(c), and $[50, 230]$ in Fig.~\ref{fig:appendix_comparison}(d).
We assume that if the results converge at $\lambda'/t$ for a certain fitting time-window, then for all $\lambda/t > \lambda'/t$ the same fitting time-window will lead to a similar convergence. Furthermore, we assume that the convergence does not strongly depend on $\beta$, as seen in Fig.~\ref{fig:appendix_comparison}.   
\section{Imbalance oscillations in the Fibonacci limit   \label{appendix_C}}
\begin{figure}[!t]
	\centering
    \includegraphics[width=\columnwidth]{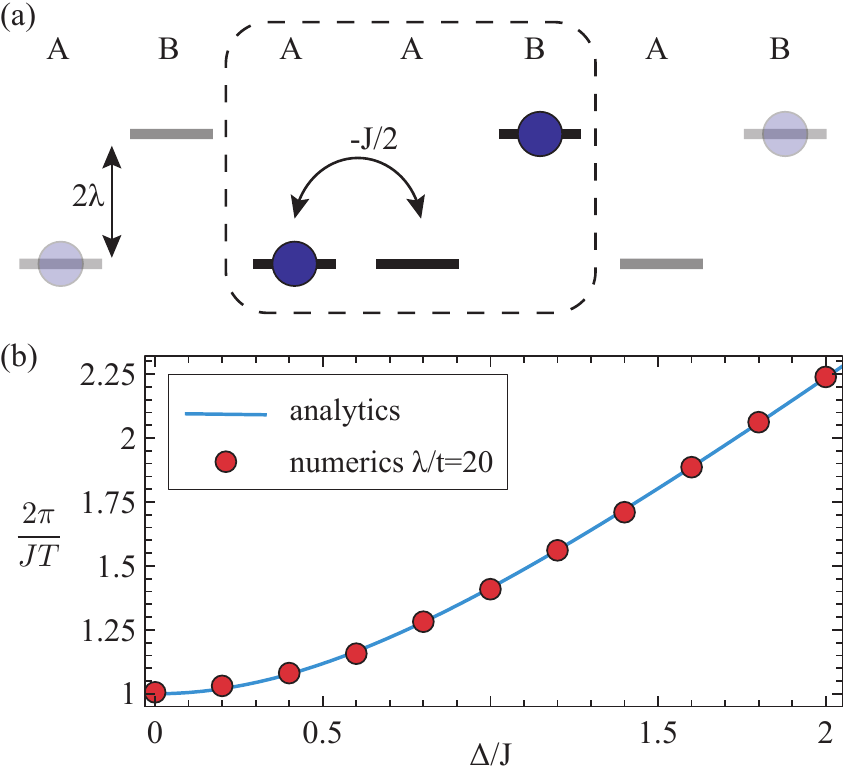}
    \caption{ Imbalance oscillations in the Fibonacci model.
    (a) 
    Sketch of the mechanism responsible for strong oscillations in the Fibonacci limit. A particle, initially placed in a well formed by two ``A'' sites (with the lower potential $-\lambda$) and bounded by ``B'' sites (with the higher potential $+\lambda$), hops inside the well, giving rise to the oscillations of the imbalance, see Figs.~\ref{fig:fibonacci}(a) and (b).
    (b)
    Comparison between the numerical result for $\lambda/t=20$ and Eq.~\eqref{eq:period_vs_D}. For the numerical calculation, we used $L=50$, $\chi=64$ and $\delta \tau=0.1$. 
    }
    \label{fig:appendix_imbalance_oscillations}
\end{figure}

Here, we describe the mechanism behind the strong oscillations of the imbalance mentioned in Section~\ref{subsection:Fibonacci}. In the Fibonacci limit, the potential modulation will contain only two values, namely $\pm \lambda$, cf.~Fig.~\ref{fig:intro}(a). We refer to a site with on-site potential $-\lambda$ ($+\lambda$) as A-site (B-site). Note that the Fibonacci sequence allows for two neighboring sites of type A, but not for two B-sites. Therefore, the combination BAAB is possible (see Fig.~\ref{fig:appendix_imbalance_oscillations}(a)), and it will occur many times throughout the chain. This configuration we dub as “a well”. In the initial N\'{e}el state, there will always be two particles inside the BAAB subsystem, one at a B-site, and the other inside the well, see Fig.~\ref{fig:appendix_imbalance_oscillations}(a).

Let us now consider a system that is deeply within the MBL phase, namely when $\lambda/t \gg 1$. The particles at isolated A-sites and at all B-sites remain localized for long times, while particles residing in their respective wells can hop from one A-site of the well to the other. To see this effect, it is sufficient to concentrate on the three sites marked with the dashed box in Fig.~\ref{fig:appendix_imbalance_oscillations}(a). Such hopping changes the value of the imbalance, and is not influenced by the value of $\lambda$.  Furthermore, because the particle at site B remains localized, we can write the Hamiltonian for a single particle inside the well, and incorporate the interaction with the particle at the B-site through a renormalized on-site energy, see Eq.~\eqref{eq:well_hamiltonian} below. The basis we choose is \{ $\ket{10}$, $\ket{01}$ \}, where the first (second) state represents the particle sitting on the left (right) A-site of the well. The Hamiltonian in such a basis reads \begin{align}
    H_\text{AA} = 
    \begin{pmatrix}
        0 & -\frac{J}{2} \\
        -\frac{J}{2} & \Delta
    \end{pmatrix} \, ,
    \label{eq:well_hamiltonian}
\end{align}
with the eigenenergies
\begin{align}
    E_{\pm} = \frac{\Delta}{2} \pm \frac{\sqrt{\Delta^2+J^2}}{2} \, .
\end{align}

Initially, the particle is placed in the left A site of the well, and the initial state is $\psi_0=\ket{10}$. 
We can now evolve the system state with time $\tau$ according to 
$$\ket{\psi(\tau)} = \exp(-i H_\text{AA} \tau)\, \ket{10},$$ 
and obtain 
\begin{align}
    \ket{\psi(\tau)} = &e^{-i\frac{\Delta}{2}\tau} \Bigg[ \cos(\sqrt{1+\xi^2} \frac{J}{2} \tau) \nonumber\\
    & \quad \qquad+ i \, \frac{\xi}{\sqrt{1+\xi^2}} \sin(\sqrt{1+\xi^2}\frac{J}{2} \tau)\Bigg] \ket{10} \nonumber\\
    &+ i\, e^{-i\frac{\Delta}{2}\tau} \frac{1}{\sqrt{1+\xi^2}}\sin(\sqrt{1+\xi^2}\frac{J}{2} \tau)\ket{01}
\end{align}
with $\xi \equiv \Delta/J$. The expectation values for the densities on the left and right A-site of the well are readily obtained as
\begin{align}
    \bra{\psi(\tau)} n^{\phantom\dagger}_{\rm left} \ket{\psi(\tau)} =& \cos^2\Big(\sqrt{1+\xi^2}\frac{J}{2} \tau\Big), \nonumber\\
    &+ \frac{\xi^2}{1+\xi^2} \sin^2\Big(\sqrt{1+\xi^2}\frac{J}{2} \tau\Big) \\
    \bra{\psi(\tau)} n^{\phantom\dagger}_{\rm right} \ket{\psi(\tau)} =&
    \frac{1}{1+\xi^2} \sin^2\Big(\sqrt{1+\xi^2}\frac{J}{2} \tau\Big) \,.
\end{align}
The particle imbalance is then calculated as
\begin{align}
    \mathcal{I}(t) &= \left(\expval{n^{\phantom\dagger}_{\rm left}} - \expval{n^{\phantom\dagger}_{\rm right}} \right)  \nonumber
    \\
    &= \frac{1}{(1+\xi^2)} \left[ \xi^2 + \cos(\sqrt{1+\xi^2} \, J \tau) \right] \, .
\end{align}

This simple exercise leads to two important conclusions. First, it follows that the period of oscillations is given by
\begin{align}
    T = \frac{2 \pi}{J \, \sqrt{1+\xi^2} } \, ,
    \label{eq:period_vs_D}
\end{align}
namely, it reduces with increasing interaction strength.
In Fig.~\ref{fig:appendix_imbalance_oscillations}(b), we show the comparison of Eq.~\eqref{eq:period_vs_D} with the period of oscillations from numerically calculated imbalance. The period obtained from numerics agrees well with the analytical prediction~\eqref{eq:period_vs_D}.
Second, the amplitude of oscillations decreases with increasing $\xi$ until they completely vanish for infinite interaction strength, i.e., $\lim_{\xi \to \infty} \mathcal{I}(t) = 1$. Such an effect can be seen in Figs.~\ref{fig:fibonacci}(a) and (b) and Fig.~\ref{fig:appendix_fibonacci}(a).

\section{Mean gap ratio   \label{appendix_D}}
To complement our conclusions in Section~\ref{subsection:i-IAAF} regarding the shrinking of the ergodic phase with increasing $\beta$ in the strongly interacting case, we use ED on shorter chains, and study the spectral gap statistics of the system. Specifically, the gap ratios measure the energy level repulsion~\cite{oganesyan2007}, and provide us with a measure for the localization in the system.
The gap ratio $r_n$ between two consecutive gaps is defined as $$0 \leq r_n \equiv \min[\delta_n, \delta_{n-1}] / \max[\delta_n, \delta_{n-1}] \leq 1,$$ where $\delta_n = E_{n+1}-E_n \geq 0$ and $n$ labels the eigenenergies from low to high.

\begin{figure}[h!]
	\centering
    \includegraphics[width=\columnwidth]{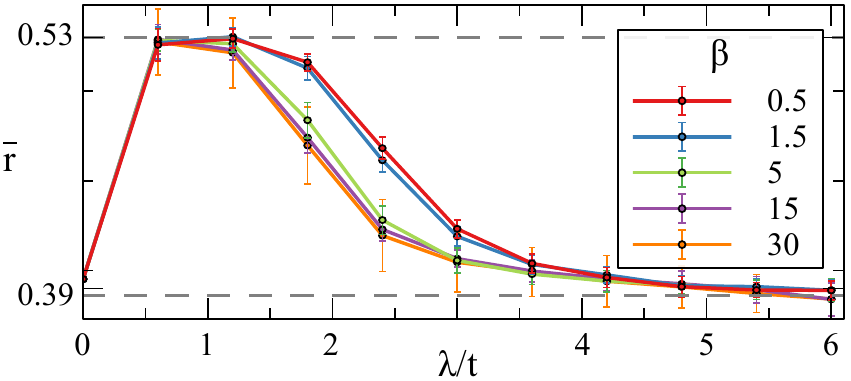}
    \caption{ 
    Mean gap ratio $\overline{r}$  as a function of $\lambda/t$ for several different values of $\beta$. We used a system with strong interaction $\Delta=1$ and $L=14$.
    }
    \label{fig:mean_gap_ratio}
\end{figure}

Taking the average over all gap ratios in the spectrum and over 36 different realizations of the quasiperiodic potential, we obtain the mean gap ratio $\overline{r}$, see Fig.~\ref{fig:mean_gap_ratio}.
Inside the ergodic phase, energy levels follow the Wigner-Dyson statistics in the thermodynamic limit, and the mean gap ratio approaches the value~\cite{atas2013} $r^{\phantom\dag}_{\rm WD} \approx 0.53$. In the MBL phase, they follow Poisson statistics and saturate at the lower value of $r^{\phantom\dag}_{\rm P} \approx 0.39$.
Although it is difficult to precisely locate the value of the critical disorder $\lambda^{\phantom\dag}_{\rm C}/t$ from $\overline{r}$ in such short chains, the overall behavior clearly shows that for higher $\beta$, all points are shifted towards smaller values of $\lambda/t$ values. This is in agreement with the TDVP results in Figs.~\ref{fig:MB_phase_diagram}(a), namely, that the ergodic phase shrinks with the increasing of $\beta$. 

\end{document}